\documentclass[a4paper,11pt]{article}
\usepackage{pos}

\title{Hide-and-seek with cosmic tau neutrinos}

\author*{Yasaman Farzan}

\affiliation{School of physics, Institute for Research in Fundamental Sciences (IPM),\\
	P.O. Box 19395-5531, Tehran, Iran}

\emailAdd{yasaman@theory.ipm.ac.ir}

\abstract{We first revisit the possibility of preserving the original flavor ratio of  high energy cosmic neutrino flux by turning on a coupling between these neutrinos and ultra-light dark matter. We discuss the bound that can be set on such a coupling from the recent $\nu_\tau$ observation by ICECUBE and  outline the implications of the coupling for the  EeV range cosmic  neutrino flux to be observed by upcoming neutrino detectors. We  then focus on the  $3+1$ scheme  when the active sterile oscillation length is of order of 1000~km for EeV range cosmic neutrinos. We show that within this scenario, the probability of survival of an active neutrino passing through the Earth can be sizable, despite the fact that the mean free path of the EeV neutrinos is much smaller than the Earth radius. This opens up the possibility to have neutrino events similar to the two anomalous events reported by ANITA.}

\FullConference{%
  *** The European Physical Society Conference on High Energy Physics (EPS-HEP2021), ***\\
  *** 26-30 July 2021 ***\\
  *** Online conference, jointly organized by Universität Hamburg and the research center DESY ***
}

\begin{document}
\maketitle

\section{Introduction}
Within the Standard Model (SM), there are three neutrinos, $\nu_e$, $\nu_\mu$ and $\nu_\tau$ which have universal couplings to the SM gauge bosons. While in recent decades the interactions of $\nu_e$, $\nu_\mu$ and their antiparticles have been studied in great detail by various experiments, the available data on $\nu_\tau$ and $\bar{\nu}_\tau$ interactions is only meager. This makes the third generation of leptons more enigmatic, raising the hope to discover new physics through  more precise study of $\nu_\tau$ and $\bar{\nu}_\tau$.
In near future, the third generation neutrinos will be pushed towards spotlight by upcoming experiments. FASER$\nu$ and SND@LHC will collect $\sim 70$ Charged Current (CC) events produced by the $\nu_\tau$ and $\bar{\nu}_\tau$ fluxes emitted from the ATLAS Interaction Point (IP) during the run III of the LHC in 2022-2024 \cite{Kling:2021gos}. The detection of cosmic neutrinos with energies higher than $\sim$PeV at neutrino telescopes opens new possibilities for studying $\nu_\tau$ and $\bar{\nu}_\tau$. Going beyond EeV, $\nu_\tau$ and $\bar{\nu}_\tau$ enjoy special privilege in detection because they can lead to Extensive Air Shower (EAS) signals to be detected by  radio  telescopes such as ANITA and by future observatories such as POEMMA \cite{Olinto:2019euf}, TRINITY \cite{Otte:2019knb} and GRAND \cite{Fang:2017mhl}. 

What makes studying high energy $\nu_\tau$ and $\bar{\nu}_\tau$ even more intriguing is the two anomalous events registered by ANITA in 2006 and 2014 \cite{ANITA:2016vrp}. The two events look like Extensive Air Shower from $\nu_\tau$ and  $\bar{\nu}_\tau$ events with energies of $0.6 \pm 0.4$ EeV and $0.56^{+0.4}_{-0.2}$ EeV emerging from $-27.4\pm 0.3^\circ$ and $-35.0\pm 0.3^\circ$ which correspond to chord sizes of 5800~km and 7300~km \cite{ANITA:2016vrp}. The reason why these events are anomalous is that at these energies, the Earth is opaque for neutrinos. 
Although some interpretation within SM, like reflection of the radio waves from layers inside the arctic ice, have been suggested \cite{Shoemaker:2019xlt,deVries:2019gzs}, the report has stirred activity among particle physicists to come up with beyond SM explanations for these two events \cite{Esmaili:2019pcy,Cherry:2018rxj}. One of the first ideas that was put forward to explain the two anomalous events was fourth sterile neutrino mixed with $\nu_\tau$. Following \cite{Farzan:2021gbx}, in this letter, we shall revisit this explanation.

\section{Flavor composition of  cosmic neutrinos with and without dark matter effects \label{sec:DM}}
Ultra high energy cosmic neutrinos can be produced by collision of very high energy protons accelerated at sources such as Active Galactic Nuclei (AGN) or gamma ray bursters on a background photon gas or on protons. For example, the scattering of cosmic ray protons off the Cosmic Microwave Background (CMB) can lead to the famous cosmogenic neutrinos with energies of EeV. 
Regardless of whether protons scatter off photons or protons, the flavor composition of neutrinos at the source will be
$F_{\nu_e}:F_{\nu_\mu}:F_{\nu_\tau}=1:2:f$ with $f\ll 1$. In case of $pp$ scatterings, there will be only a small $\nu_\tau$ flux from the decay of $D$ mesons that are produced via Charged Current (CC) interaction of the $s$ partons. In case of $p \gamma$ scattering, the $\nu_\tau$ component of the flux at the source will be even smaller and suppressed by intrinsic $c$ parton density. Nevertheless,  the neutrino propagation from the cosmic sources to the Earth will lead to neutrino oscillation and therefore a democratic flavor composition when they reach the Earth: $F_{\nu_e}^\oplus :F_{\nu_\mu}^\oplus:F_{\nu_\tau}^\oplus=1:1:1$. This prediction turns out to be quite robust against varying the conditions at the source. Even by invoking new physics, it is not so trivial to obtain an arbitrary flavor composition at the Earth. It has been shown in \cite{Bustamante:2015waa} that as long as the flux arriving at the Earth is an arbitrary incoherent composition of the neutrino mass eigenstates, the deviation from the $1:1:1$ prediction will be small. In particular, the $\nu_\tau$ flux at the Earth will be nonzero.

 It has been shown in \cite{Farzan:2018pnk} that if neutrinos couple to the background of ultralight DM, the original flavor ratio of neutrinos from a source located in a DM halo can be maintained. That is even at the Earth, we expect vanishing $\nu_\tau$ and $\bar{\nu}_\tau$  fluxes, $F_{\nu_e}:F_{\nu_\mu}:F_{\nu_\tau}=1:2:0$. 
 The reason is that the DM background induces an effective flavor diagonal mass for neutrinos which in regions where DM density is relatively high  can dominate over the mass term (more precisely, DM induces effective mass$\gg \Delta m^2/E_\nu$). As a result, in  the DM halo of the galaxies, the energy and flavor eigenstates coincide so there will be no oscillation.  Since the variation of the DM density along the route of neutrinos from the source to the Earth is smooth, the flavor transition will be adiabatic. Thus, if both source and detector are inside some DM halos, the original flavor ratios will not change. 
 As discussed in \cite{Farzan:2021gbx}, the observation of the  two $\nu_\tau$ events by ICECUBE \cite{IceCube:2020abv}
 with energies of O(PeV) constrains this scenario, putting an upper bound on the relevant coupling. Saturating this bound, it is still possible for EeV neutrinos to maintain their original flavor composition provided that they  originate in a region with relatively high DM density. The preservation of the original flavor ratio does not apply for neutrinos coming from regions outside the galactic DM halos where the density of DM is at the level of the average DM density in the Universe \cite{Farzan:2021gbx}. As a result, even if  DM induced effective mass in our halo is large, the flavor of cosmogenic neutrinos arriving at the Earth will still be the canonical prediction: $F_{\nu_e}^\oplus:F_{\nu_\mu}^\oplus:F_{\nu_\tau}^\oplus=1:1:1$.

The ARA and ARIANNA detectors are sensitive to Askaryan emission from all three flavors. On the other hand, detectors such as GRAND can measure the flux of Earth skimming $\nu_\tau$. As discussed in \cite{Farzan:2021gbx}, observation of $\nu$ flux through Askaryan effect along with null result on  the $\nu_\tau$ component will hint towards dark matter effect. Moreover, it will be an indication that the flux is not of cosmogenic origin but the source is located in a DM halo.

At energies of EeV, the Earth is opaque for neutrinos. The $\nu_e$ and $\nu_\mu$ fluxes with such energies will be completely absorbed but the $\nu_\tau$ flux can go through regeneration by producing $\tau$ via scattering on the nuclei and the subsequent $\tau$ decay into $\nu_\tau$ with lower energy.
An EeV $\nu_\tau$ flux entering the Earth from one side will lead to a flux of  PeV energy neutrino flux detectable by ICECUBE. From this consideration, a bound can be obtained on the flux of EeV $\nu_\tau$ flux arriving at the Earth from the lack of detection of a lower energy counterpart by ICECUBE.  Indeed, Ref.~\cite{Safa:2019ege} shows that the $\nu_\tau$ flux required to explain the two anomalous events reported by ANITA overshoots the bound by more than seven orders of magnitudes.
\section{Ultra high energy neutrino flux within the 3+1 neutrino scheme\label{3+1}}
Let us now discuss the propagation of ultra high energy 
cosmic neutrinos within the 3+1 scheme. In general, the neutrino flavor states $ (\nu_e,\nu_\mu,\nu_\tau,\nu_s)$ are related to neutrino mass eigenstates $(\nu_1,\nu_2,\nu_3,\nu_4)$ via a unitary $4\times 4$ matrix that can be denoted by $U$. In the flavor basis, the evolution of neutrinos (in absence of dark matter effects) are governed by the following relation 
 \begin{eqnarray} i \frac{d \psi}{dt}&= &\left[U\cdot {\rm diag} (0,\Delta m_{21}^2/(2E_\nu),\Delta m_{31}^2/(2E_\nu), \Delta M^2/(2E_\nu) )\cdot U^\dagger\right.\cr &+& {\rm diag}(\sqrt{2} G_F(N_e-N_n/2),-\sqrt{2} G_FN_n/2,-\sqrt{2} G_FN_n/2,0)\cr
 &-& \left. i ~{\rm diag}(\Gamma/2,\Gamma/2,\Gamma/2,0)\right]\psi \end{eqnarray}
 where $\Delta M^2$ denotes the mass splitting of the fourth neutrino ($\Delta M^2\equiv m_4^2-m_1^2$) and $\Gamma$ in the last line takes care of scattering of active neutrinos off the nuclei. The evolution of antineutrinos are given by similar equation replacing $U\to U^*$ and changing the sign of the matter effect in the second line.
 The evolution in presence of $\Gamma$ is non-unitary simply because we have not included the charged particles produced by $\nu$ interaction in our analysis. That is the system under consideration is not closed.
 From this relation, it is obvious that in the limit $\Gamma L\gg 1$ and $\Delta M^2L/2E_\nu\ll 1$, the propagation will remove the active components of  a mass eigenstate. That is $|\nu_i\rangle \to U_{si}^*|\nu_s\rangle$.  As a result, the notion that $|\nu_4\rangle$  propagating in the matter survives longer than active neutrinos is correct because the cross section is suppressed by $1-|U_{s 4}|^2$ but the active components of $\nu_4$ becomes absorbed. Thus, the survived components, being sterile, cannot lead to charged current interactions. As a result, contrary to the claims in  the literature, explaining the ANITA anomalous events by $\nu_4$ with a mean free path of  the size of the Earth radius $\Gamma R_\oplus \sim 1$ is not  so trivial. In the following, we examine this solution for $\Delta M^2 R_\oplus/(2E_\nu)\sim 1 $.

  \begin{figure}[h]
  	\hspace{3 cm}
  	\includegraphics[width=0.60\textwidth, height=0.4\textwidth]{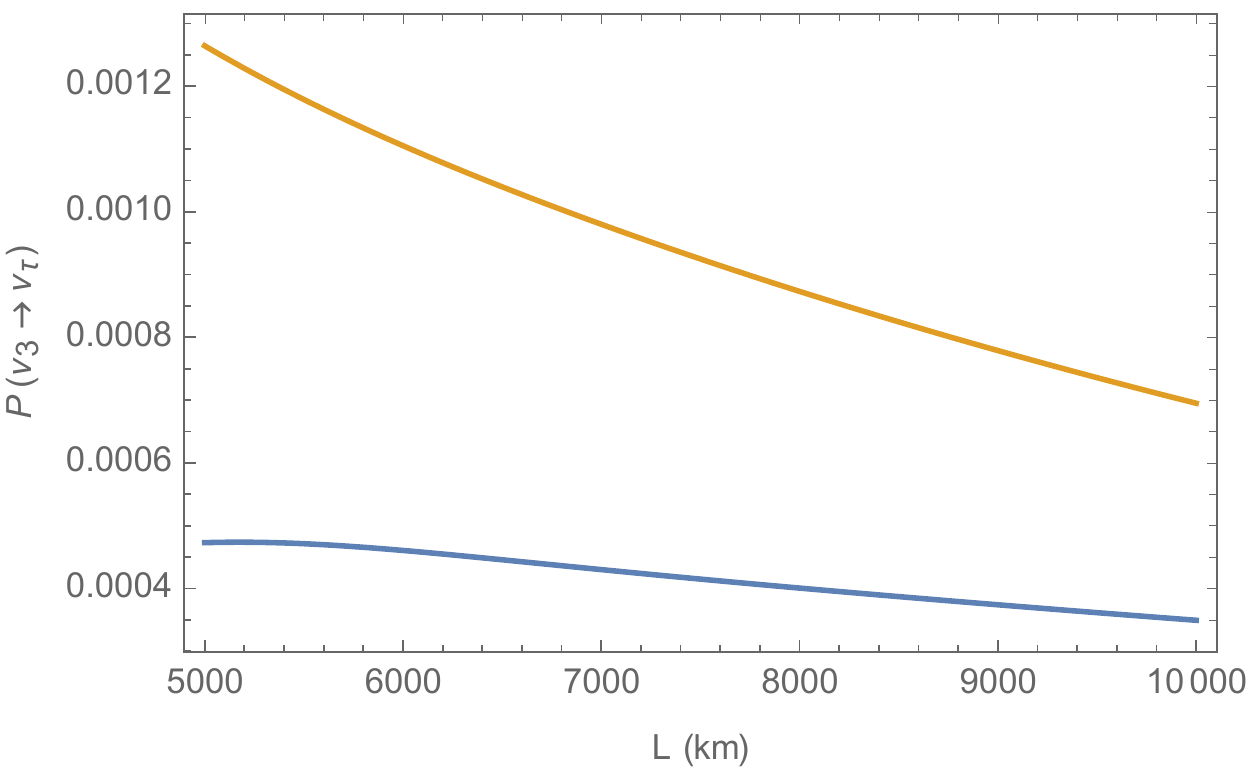}
  	\caption[...]{ Probability  $\nu_3 \to \nu_\tau$ and $\bar{\nu}_3 \to \bar{\nu}_\tau$  versus the size of the traversed chord.  We have taken $|U_{\tau 4}|^2=0.1$, $|U_{e4}|^2=|U_{\mu 4}|^2=0$, $\Gamma=2.96\times 10^{-3}$ km$^{-1}$ and $\Delta M^2/E_\nu= 10^{-3}$ km$^{-1}$.  The   blue  and orange lines respectively correspond to the   neutrino and anti-neutrino modes.
  	}
  	\label{fig:3+1}
  \end{figure}
  Unless there is a new mechanism for sterile neutrino production, the flavor composition of neutrinos at the source from meson (pion and Kaon) decay will be $(F_{\nu_e}:F_{\nu_\mu}:F_{\nu_\tau}:F_{\nu_s})\simeq (1:2:0:0)$ which in the mass basis can be written as
  $$(F_{\nu_1}:F_{\nu_2}:F_{\nu_3}:F_{\nu_4})\simeq
  (|U_{e1}|^2+2|U_{\mu 1}|^2:|U_{e2}|^2+2|U_{\mu 2}|^2:|U_{e3}|^2+2|U_{\mu 3}|^2:|U_{e4}|^2+2|U_{\mu 4}|^2).$$
  In particular,  if $\nu_s$ only mixes with $\nu_\tau$ (as the model in \cite{Cherry:2018rxj}), $U_{e 4}=U_{\mu 4}=0$ so the $\nu_4$ component vanishes.
  In absence of the DM induced effective mass, the propagation of the long distance between the source and the detector will lead to decoherence of mass eigenstates.  Once these mass eigenstates with energies of EeV enter the Earth, their active components become eventually absorbed. On the other hand, the remaining sterile component can oscillate back to the active ones, thanks to the $\Delta M^2$ splitting. Fig.~\ref{fig:3+1} shows $P(\nu_3 \to \nu_4)$ and $P(\bar{\nu}_3 \to \bar{\nu}_4)$ versus the traversed chord size $L$ taking $|U_{e4}|=|U_{\mu 4}|=0$ but $|U_{\tau 4}|=0.1$. Such large value of  $|U_{\tau 4}|$ is still allowed \cite{Dentler:2018sju}. The value of
  $\Gamma=0.00296$ km$^{-1}$ corresponds to the rate of neutrino absorption in 
  the mantle for neutrinos of EeV energy \cite{Farzan:2021gbx}.  The oscillation length between $\nu_s$ and $\nu_\tau$ is taken equal to 1000~km.  As seen from the Fig.~\ref{fig:3+1}, the oscillation probability can be as large as  0.0012. This probability should be compared to the   survival  probability of $\nu$  in the absence of $\nu_4$, $e^{-\Gamma L}=10^{-7}-10^{-12}$. Thus, the $3+1$ scheme enhances the probability of $\nu_\tau$ emerging after traversing chords with sizes  larger than 5000~km by a factor larger than $10^4-10^9$. Such an enhancement raises the hope to see $\nu_\tau$-like events by future radio detectors such as POEMMA or GRAND. As mentioned before, Ref.~\cite{Safa:2019ege} sets an upper bound on the flux of EeV neutrinos entering the Earth.  Saturating this bound,  a radio detector with an acceptance of $\sim 10^{10}$ cm$^2$ ({\it i.e.,} one order of magnitude above that of ANITA)
can see handful of $\nu_\tau$ events from chords that completely absorb the neutrino flux within the $3 \nu$ scheme.  
\section{Summary and discussion}

  We have reviewed two scenarios that can affect the cosmic high energy $\nu_\tau$ flux: (i) effective neutrino mass induced by background ultra-light dark matter; (ii) the $3+1$ scheme with $\nu_s$ mixed with $\nu_\tau$. We argued that if both the source and the detector of high energy cosmic neutrinos are located in the dark matter halos, the dark matter induced effective mass can dominate over the Hamiltonian in the vacuum and therefore the original flavor composition of neutrinos can be maintained. This means that if flavor ratio at the source is $(F_{\nu_e}:F_{\nu_\mu}:F_{\nu_\tau})=(1:2:0)$ (as predicted in the canonic scenario), the flux at the detectors will not have the $\nu_\tau$ component. Observation of the two $\nu_\tau$ events at ICECUBE sets an upper bound on the effective coupling between neutrinos and dark matter. Even satisfying this bound, the dark matter effects can maintain the original  flavor ratio of EeV neutrinos coming from a source located in a dark matter halo.
  In future, if detectors such as ARA or ARIANNA detect a flux of EeV neutrinos via their Askaryan  emission but POEMMA \cite{Olinto:2019euf}, TRINITY \cite{Otte:2019knb} and GRAND \cite{Fang:2017mhl} do not detect matching $\nu_\tau$ component, it will be very challenging to find an explanation within the standard picture or even within the famous beyond SM scenarios such as neutrino decay or sterile neutrino. In our scenario, this observation can however easily be attributed  to the production of the neutrino in a source immersed in a halo of ultra-light dark matter with flavor-conserving couplings to the  neutrinos. It will also implicitly mean the observed EeV neutrino flux  is not of cosmogenic origin. 
  
  We then discussed the propagation of EeV neutrinos inside the Earth within the $3+1$ scheme with an oscillation length of 1000 km. We showed that active sterile oscillation will enhance the probability of the survival of an active component crossing the Earth by a factor of $10^4-10^9$ relative to the standard picture where neutrinos of EeV energy become absorbed inside the Earth. This increases the hope to observe EeV neutrinos emerging from deep down the Earth.

\acknowledgments
This letter is prepared  for proceedings of EPS-HEP 2021. The author would like to thank the organizers of this event.
This project has received funding /support from the European Union$^\prime$s Horizon 2020 research and innovation programme under the Marie Sklodowska -Curie grant agreement No 860881-HIDDeN. The author has received  financial support from Saramadan under contract No.~ISEF/M/400279 and No.~ISEF/M/99169.
  

\end{document}